\documentstyle[11pt]{article}
\input epsf
\epsfverbosetrue
\begin{document}  
\renewcommand{\thefootnote}{\alph{footnote}}
\begin{titlepage}
\begin{tabbing}
\hspace{11cm} \= HIP -- 1999 -- 21 / TH \\
\> \today
\end{tabbing}

\begin{center}
\vfill
{ \Large\bf A coupled $K$-matrix description of the reactions $\pi 
N\rightarrow \pi  N$, $\pi  N\rightarrow \eta N$,
$\gamma N\rightarrow \pi N$ and
$\gamma N\rightarrow \eta N$}

\vspace{0.5cm}
A.M.~Green\footnotemark[1]$^,$\footnotemark[3],
 and S.~Wycech\footnotemark[2]$^,$\footnotemark[4] \\

$\,^a$Department of Physics and Helsinki
Institute of Physics, P.O. Box 9,\\
FIN--00014 University of Helsinki, Finland \\
$\,^b$Soltan Institute for Nuclear Studies, Warsaw, Poland
\end{center}
\setcounter{footnote}{3}
\footnotetext{email: {\tt anthony.green@helsinki.fi}}
\setcounter{footnote}{4}
\footnotetext{email: {\tt wycech@fuw.edu.pl}}
\setcounter{footnote}{0}

\date{\today}

\begin{abstract}
The coupled $\pi N$, $\eta N$ , $\gamma N$ systems are described by a
 $K$-matrix  method.
The parameters in this model are adjusted to get an optimal
fit to $\pi  N\rightarrow \pi  N$, $\pi  N\rightarrow \eta N$,
$\gamma N\rightarrow \pi N$ and
$\gamma N\rightarrow \eta N$ data in an energy range of about 100 MeV
each
side of the $\eta$ threshold. \\
The coupling of photons to the $N(1535)$ state is  extracted and also an
alternative to the current $S11\gamma N\rightarrow \pi N$ amplitudes 
suggested. Expansions are given for the $\eta \eta$ and $\gamma \eta$
amplitudes in terms of the $\eta$ momentum.
Effects of interference of this state with background potential 
interactions are discussed and experimental consequences are 
indicated. 
\end{abstract}
\end{titlepage}

\section{Introduction}
There is  an increasing interest in $\eta$-meson physics both 
experimentally and theoretically. On the experimental side several
facilities are now able to produce sufficient $\eta$'s to 
enable a study to be made of their interactions with other 
particles. In particular, the photon machines MAMI\cite{Krusche} 
and GRAAL\cite{GRAAL} are supplementing the earlier hadronic machines 
such as  SATURNE\cite{saturne},  CELSIUS\cite{uppsala} and
COSY\cite{hadronic}.
The current theoretical interest stems  partly
from  the early indications that the $\eta-N$ interaction is 
attractive and so could possibly lead to $\eta$-nucleus 
quasi-bound states (e.g. Refs.~\cite{bha},~\cite{hai}).
The theoretical
approaches fall into two main categories. In the one, the various
processes involving $\eta$-meson interactions are described in terms of
microscopic models containing baryon resonances and the exchange of
different mesons (e.g. Refs.~\cite{sau}, ~\cite{ben}) which may 
be based on a chiral perturbation approach (e.g.Ref.~\cite{kai}) or 
a quark model (e.g. Ref.~\cite{ari}).
Unfortunately, this approach requires a
knowledge of the magnitudes and relative phases of many hadron-hadron
couplings several of which are very poorly known.
In addition, since $\eta$ interactions -- in the absence of $\eta$-meson
beams -- can only be studied as final state interactions,
one has to exploit relationships between the many processes involved.
For example, in the present note, the main interest is in
the reaction a) $\gamma N\rightarrow \eta N$. However, this is dependent
on the final state interaction b) $\eta N\rightarrow \eta N$, which
in turn depends on the reactions c) $\pi  N\rightarrow \eta N$ and d)
$\pi  N\rightarrow \pi  N$. Similarly, reactions c) and d) are related to 
e) $\gamma N\rightarrow \pi N$. Therefore, any model that claims to
describe reaction a) must also see its implications in reactions
b), .., e). This, we believe, is too ambitious a program at present. 
At this stage it is probably more informative to check the consistency
between the data of the above five reactions and be able to relate them
in terms of a few phenomenological parameters.
When this has been accomplished, it will hopefully be possible to
understand these parameters in terms of more microscopic models.
With this in mind, in Ref.~\cite{GW97} a $K$-matrix model was developed
by the authors to describe the reactions a), b), c) and d)
in an energy range  of about 100 MeV
each side of the $\eta$ threshold. This model was expressed in the form
of two coupled channels for  $s$-wave $\pi-N$ and  $\eta-N$ scattering
with the effect of the two pion channel ($\pi  N\rightarrow \pi \pi N$) 
being included only implicitly. 
The latter was achieved by first introducing the two pion
process as a third channel in the $K$-matrix and subsequently
eliminating that channel  as an "optical potential" correction to the 
other two channels. It should be emphasized that this is not an
approximation but is done only for convenience, since we do not address 
cross sections involving explicitly two final state pions.

In Ref.~\cite{GW97} the $\eta$-photoproduction cross section was assumed to
be proportional to the elastic $\eta-N$ cross section
($|T_{\eta\eta}|^2$). This is in line with the so-called Watson 
approximation \cite{watson}.
 In this way each of the matrix elements in  the two-by-two $T$-matrix 
of Ref.~\cite{GW97} was associated with some specific experimental data --
$T_{\pi \pi}$ with the $\pi N$ amplitudes of Arndt {\em et al.}
\cite{Arndt}, $T_{\pi \eta}$ with the $\eta$-production cross section in
the review by Nefkens\cite{Nefkens} and $T_{\eta \eta}$ with the 
$\eta$-photoproduction cross section of Krusche {\em et al.}\cite{Krusche}.

In this note we now wish to treat the $\gamma N$ channel  explicitly. 
An enlargement
of the $K$-matrix basis then permits a direct estimate of the matrix element 
$T_{\gamma \eta}$, so that 
$\sigma (\gamma N\rightarrow \eta N) \propto |T_{\gamma\eta}|^2$, thereby
avoiding the earlier
 assumption that $\sigma (\gamma N\rightarrow \eta N) \propto |T_{\eta\eta}|^2$.
The $K$-matrix would now be a four-by-four matrix with the
channels $\pi N$, $\eta N$, $\pi \pi N$ and $\gamma N$. In
principle,  10 different processes, corresponding to each matrix element,
could be analysed simultaneously. However, in
practice, it is more convenient to elimate some channels by the 
"optical potential" method used already in Ref.~\cite{GW97}. We,
therefore, describe in Section 2 the above reactions in terms of three 
separate $T$-matrices. 
In Section 3, we give the fitting strategy and also the numerical results in
terms of the 13 parameters needed to specify the $K$-matrices. This
section also includes expansions -- in terms of the $\eta$ momentum -- for
the amplitudes of the $\eta N\rightarrow \eta N$ and 
$\gamma N\rightarrow \eta N$ reactions near the $\eta$ threshold.
Section 4 contains a discussion and some conclusions.

\section{The $K$-matrix formalism}
In principle, the four channels of interest  -- $\pi N$, $\eta N$, 
$\pi \pi N$ and $\gamma N$ -- should be treated simultaneously.
However, it is more convenient and transparent if the problem is
analysed in terms of three separate $T$-matrices.

\subsection{Coupled $\pi N$ and  $\eta N$ channels}

 The first $T$-matrix is precisely the same as in Ref.~\cite{GW97}, where only 
the $\pi N$
and  $\eta N$ channels -- denoted by the indices $\pi$, $\eta$ -- are explicit. This
can be written as 
\begin{equation}
\label{KT1}
  T_1 = \left( \begin{array}{cc}
T_{\pi \pi}&T_{\pi \eta}\\
T_{ \eta\pi}&T_{ \eta\eta}
\end{array} \right)
= \left( \begin{array}{ll}
 \frac{A_{\pi \pi}}{1-iq_{\pi}A_{\pi \pi}}&  
\frac{A_{\pi \eta}}{1-iq_{\eta}A_{\eta \eta}}\\
  \frac{A_{\eta \pi}}{1-iq_{\eta}A_{\eta \eta}}
&  \frac{A_{\eta \eta}}{1-iq_{\eta}A_{\eta \eta}}\end{array} \right),
\end{equation}
where $q_{\pi,\eta}$ are the center-of-mass momenta of the two mesons in the two 
channels $\pi,\eta$ and
the channel scattering lengths $A_{ij}$ are expressed in terms of the 
$K$-matrix elements, via the solution of $T=K+iKqT$, as
\begin{center}
$A_{\pi \pi}=K_{\pi \pi}+iK^2_{\pi \eta}q_{\eta}/(1-iq_{\eta}K_{\eta \eta})$,
 \ \ $ A_{\eta \pi}= A_{ \pi \eta}=K_{\eta \pi}/(1-iq_{\pi}K_{\pi \pi}) $
\end{center}
\begin{equation}
\label{2.2}                                 
A_{\eta \eta}=K_{\eta \eta}+iK^2_{\eta \pi}q_{\pi}/(1-iq_{\pi}K_{\pi \pi}).
\end{equation}
At this stage the $\pi\pi N$ channel is incorporated as an "optical model"
correction to the corresponding matrix element of $T_1$
and the $\gamma N$ channel is simply ignored since
this $T$-matrix is used to describe only reactions b), c) and d), where the
effect of the $\gamma N$ channel is small being  only an electromagnetic
correction to these three reactions. As discussed in Ref.~\cite{GW97} various features of the 
experimental data suggest that the $K$-matrix elements can be
parametrized in terms of energy independent constants -- the background
terms $B_{i j}$ -- plus poles associated with the $S$-wave $\pi N$
resonances $N(1535)$ and $N(1650)$. This results in 
 \begin{center}
$ K_{\pi\pi}\rightarrow K_{\pi\pi}(a)=\frac{\gamma_{\pi}(0)}{E_0-E}+
\frac{\gamma_{\pi}(1)}{E_1-E}+i\frac{K_{\pi 3}q_3K_{3 \pi}}{1-iq_3K_{33}}$ ,
 \ \ $ K_{\pi\eta}\rightarrow  B_{\pi\eta}+
\frac{\sqrt{\gamma_{\pi}(0)\gamma_{\eta}(0)}}
{E_0-E}+i\frac{K_{\pi 3}q_3K_{3 \eta}}{1-iq_3K_{33}},$
\end{center}
\begin{equation}
\label{Ks}
 K_{\eta\eta}\rightarrow K_{\eta\eta}(a)=B_{\eta\eta}+
\frac{\gamma_{\eta}(0)}{E_0-E}+i\frac{K_{\eta 3}q_3K_{3 \eta}}{1-iq_3K_{33}},
\end{equation}
where
\[ K_{33}= \frac{\gamma_3(0)}{E_0-E}+\frac{\gamma_3(1)}{E_1-E} ,
\ \ \ \ \  K_{\pi 3}= \frac{\sqrt{\gamma_{\pi}(0)\gamma_{3}(0)}}{E_0-E} 
+\frac{\sqrt{\gamma_{\pi}(1)\gamma_{3}(1)}}{E_1-E}, \ \ \ 
 K_{\eta 3}= \frac{\sqrt{\gamma_{\eta}(0)\gamma_{3}(0)}}{E_0-E}.\]
The last terms on the RHS of Eqs.~(\ref{Ks}) represent the effect of the
eliminated $\pi \pi N$ channel.

\vskip 0.4cm

\subsection{Coupled $\eta N$ and  $\gamma N$ channels}
 The second $T$-matrix involves only the two channels $\eta N$ and 
$\gamma N$  -- denoted by the indices  $\eta$,$\gamma$  -- where now it is  the  
$\pi \pi N$ and $\pi N$ channels that are treated as optical potentials. 
This $T$-matrix is written as  
\begin{equation}
\label{KT2}
   T_2 = \left( \begin{array}{cc}
T_{\eta \eta}&T_{\gamma \eta}\\
T_{ \eta\gamma}&T_{ \gamma \gamma}
\end{array} \right)
=
\left( \begin{array}{cc}
 \frac{A_{\eta \eta}}{1-iq_{\eta}A_{\eta \eta}}&  
\frac{A_{\gamma\eta  }}{1-iq_{\eta}A_{\eta \eta}}\\
  \frac{A_{ \eta  \gamma}}{1-iq_{\eta}A_{\eta \eta}}
&  T_{\gamma \gamma}\end{array} \right),
\end{equation}
\[ {\rm where} \ \  
A_{\gamma \eta}=A_{\eta \gamma}=
K_{\gamma \eta}/(1-iq_{\gamma}K_{\gamma \gamma}), \ \
A_{\eta \eta}=K_{\eta \eta}+iK^2_{\gamma \eta}q_{\gamma}/(1-iq_{\gamma}
K_{\gamma \gamma}).\]
Here we are not interested in $T_{\gamma \gamma}$, since this would
describe the $\gamma N  \rightarrow \gamma N $ reaction. The forms of 
$K_{\pi\pi}(a)$, $K_{\pi\eta}$, $K_{33}$, $K_{\pi 3}$ and $K_{\eta 3}$ are
the same as given above. However,
\begin{equation}
\label{ketaeta}
 K_{\eta\eta}\rightarrow K_{\eta\eta}(b)=K_{\eta\eta}(a)+
i\frac{K_{\eta\pi} q_{\pi}K_{\pi\eta}}{1-iq_{\pi} K_{\pi\pi}(a)}.
\end{equation} 
Also we now need
\begin{equation}
\label{ketagamma}
K_{\gamma\eta}=B_{\gamma\eta}+\frac{\sqrt{\gamma_{\gamma}(0)\gamma_{\eta}(0)}}
{E_0-E}+i\frac{K_{\gamma \pi}q_{\pi}K_{\pi\eta}}{1-iq_{\pi}K_{\pi \pi}(a)}
+i\frac{K_{\gamma 3}q_{3}K_{3 \eta}}{1-iq_{3}K_{3 3}},
\end{equation}
\begin{equation}
\label{kgammagamma}
K_{\gamma \gamma}=\frac{\gamma_{\gamma}(0)}{E_0-E}+
\frac{\gamma_{\gamma}(1)}{E_1-E}+
i\frac{K_{\gamma \pi}q_{\pi}K_{\pi \gamma }}{1-iq_{\pi}K_{\pi \pi}(a)}
+i\frac{K_{\gamma 3}q_{3}K_{3 \gamma}}{1-iq_{3}K_{3 3}}
\end{equation}
and
\begin{equation}
\label{kgammapi}
K_{\gamma \pi}=B_{\gamma \pi}+
\frac{\sqrt{\gamma_{\gamma}(0)\gamma_{\pi}(0)}}{E_0-E}+
\frac{\sqrt{\gamma_{\gamma}(1)\gamma_{\pi}(1)}}{E_1-E}
+i\frac{K_{\gamma 3}q_{3}K_{3 \pi}}{1-iq_{3}K_{3 3}}
\end{equation}
 where the last terms on the RHS represent the effect of the eliminated
$\pi N$- and $\pi\pi N$-channels. Also we need
\begin{equation}
\label{kgamma3}
K_{\gamma 3}=
\frac{\sqrt{\gamma_{\gamma}(0)\gamma_{3}(0)}}{E_0-E}+
\frac{\sqrt{\gamma_{\gamma}(1)\gamma_{3}(1)}}{E_1-E}.
\end{equation}

\vskip 0.4cm
\subsection{Coupled $\pi N$ and  $\gamma N$ channels}
 The third $T$-matrix involves only the two channels $\pi N$ and 
$\gamma N$  -- denoted by the indices  $\pi$,$\gamma$  -- where now it is  the  
$\eta N$ and $\pi \pi N$   channels that are treated as optical potentials. 
This $T$-matrix is written as  
\begin{equation}
\label{KT3}
   T_3 = \left( \begin{array}{cc}
T_{\pi \pi}&T_{\gamma \pi}\\
T_{ \pi\gamma}&T_{ \gamma \gamma}
\end{array} \right)
=
\left( \begin{array}{cc}
 \frac{A_{\pi \pi}}{1-iq_{\pi}A_{\pi\pi}}&  
\frac{A_{\gamma\pi  }}{1-iq_{\pi}A_{\pi \pi}}\\
  \frac{A_{ \pi  \gamma}}{1-iq_{\pi}A_{\pi \pi}}
&  T_{\gamma \gamma}\end{array} \right),
\end{equation}
\[ {\rm where} \ \  
A_{\gamma \pi}=A_{\pi \gamma}=
K_{\gamma \pi}/(1-iq_{\gamma}K_{\gamma \gamma}), \ \
A_{\pi \pi}=K_{\pi \pi}+iK^2_{\gamma \pi}q_{\gamma}/(1-iq_{\gamma}
K_{\gamma \gamma}).\]
As before, we are not interested in $T_{\gamma \gamma}$. The forms of 
$K_{\eta\eta}$=$K_{\eta\eta}(a)$, $K_{\pi\eta}$, $K_{33}$, $K_{\pi 3}$ and $K_{\eta 3}$ are
the same as given above. However,
\begin{equation}
\label{pipi}
 K_{\pi\pi}\rightarrow K_{\pi\pi}(b)=K_{\pi\pi}(a)+
i\frac{K_{\pi\eta} q_{\eta}K_{\eta\pi}}{1-iq_{\eta} K_{\eta\eta}(a)}.
\end{equation} 
Also we now need
\begin{equation}
\label{pigamma2}
K_{\gamma\pi}=B_{\gamma\pi}+\frac{\sqrt{\gamma_{\gamma}(0)\gamma_{\pi}(0)}}
{E_0-E}+\frac{\sqrt{\gamma_{\gamma}(1)\gamma_{\pi}(1)}}
{E_1-E}
+i\frac{K_{\gamma \eta}q_{\eta}K_{\eta\pi}}{1-iq_{\eta}K_{\eta \eta}(a)}
+i\frac{K_{\gamma 3}q_{3}K_{3 \pi}}{1-iq_{3}K_{3 3}},
\end{equation}
\begin{equation}
\label{kgammagamma2}
K_{\gamma \gamma}=\frac{\gamma_{\gamma}(0)}{E_0-E}+
\frac{\gamma_{\gamma}(1)}{E_1-E}+
i\frac{K_{\gamma \eta}q_{\eta}K_{\eta \gamma }}{1-iq_{\eta}K_{\eta \eta}(a)}
+i\frac{K_{\gamma 3}q_{3}K_{3 \gamma}}{1-iq_{3}K_{3 3}}
\end{equation}
 where the last terms on the RHS represent the effect of the eliminated
$\eta N$- and $\pi\pi N$-channels. Also we need
\begin{equation}
\label{kgammapi2}
K_{\gamma \eta}=B_{\gamma \eta}+
\frac{\sqrt{\gamma_{\gamma}(0)\gamma_{\eta}(0)}}{E_0-E}
+i\frac{K_{\gamma 3}q_{3}K_{3 \eta}}{1-iq_{3}K_{3 3}}.
\end{equation}
The definitions of all other parameters are the same as  for 
$T_{1,2}$.

\vskip 0.4cm
\section{Fitting strategy and results}
Compared with Ref.~\cite{GW97} there are now four new parameters
$B_{\gamma \pi}$, $B_{\gamma \eta}$, $\gamma_{\gamma}(0)$ and  
$\gamma_{\gamma}(1)$ 
 explicitly dependent on  the index $\gamma$.  
These four parameters 
replace the single free parameter $A(Phot)$ that related
$\sigma (\gamma N\rightarrow \eta N)$ and $T_{\eta\eta}$. In all there
are now 13 parameters that are determined by a Minuit fit of upto 158
pieces of data -- 23 are $\pi N$ amplitudes (real and imaginary)\cite{Arndt}, 
11 are $\pi N\rightarrow \eta N$ cross sections[$\sigma(\pi\eta)$]
\cite{Nefkens} and  53 are $\gamma N\rightarrow \eta N$ cross sections
[$\sigma(\gamma\eta)$]\cite{Krusche}. 
In addition, from Ref.~\cite{gpivip} we use  upto 48 
$S11(\gamma N\rightarrow \pi N)$ amplitudes  in the energy
range  $1350\leq E_{c.m.} \leq 1650$MeV. There are several reasons for
choosing this  upper limit:\\
a) We wish to include the full effect of the $N(1535)$.\\
b) The $\gamma N\rightarrow \pi N$ and $\gamma N\rightarrow \eta N$
reactions are closely related and so attempting to fit them simultaneously
over very different energy ranges could give misleading results.
Therefore, we do not attempt to use the available data at higher energies.\\
c) The values of the $\gamma N\rightarrow \pi N$ amplitudes are far
from being unique  -- as is clear when comparing the amplitudes
of Refs.~\cite{gpivip} and \cite{Mainz}.  In fact, in view of this lack 
of uniqueness we do not use the quoted errors of 
Ref.~\cite{gpivip}. Instead, we make two overall fits where, in the one case,
all the errors in Ref.~\cite{gpivip} are increased to  $\pm \sqrt{2}$
 for both the Real and  Imaginary components and, in the second case, the
increase is only to $\pm 1/\sqrt{2}$. These choices were made so that the
resultant $\chi^2$/dpt for this reaction are comparable to those in the other
reactions. We realise that this procedure is throwing away information. 
However,  the main aim in this work is to
study the $\gamma N\rightarrow \eta N$ reaction with the
$\gamma N\rightarrow \pi N$ playing only a secondary role as a possible
stabilizing effect. Therefore, we  want a $K$-matrix fit that is good
for the well established reactions but, at the same time, also reproduces 
the  qualitative trends in the 
$\gamma N\rightarrow \pi N$ reaction suggested by Refs.~\cite{gpivip}
and \cite{Mainz}. In fact, we could even turn the argument around and
say that our $S11$ amplitudes are a {\em prediction} that is consistent
with the other reactions.

In practice, the actual $\eta$-production cross section data was  used in a
reduced form, from which threshold factors have been removed -- namely:
\begin{equation}
\sigma(\pi\eta)_r=\sigma(\pi\eta)\frac{q_{\pi}}{q_{\eta}}= 
\frac{8\pi q_{\pi}}{3q_{\eta}}|T_{\pi \eta}|^2\ \ \ \ 
{\rm and} \ \ \ \                                           
\tau(\gamma\eta)_r=\sqrt{\sigma(\gamma\eta)\frac{E_{\gamma}}{4\pi q_{\eta}}}
=|T_{\gamma\eta}|.
\end{equation}
In Ref.~\cite{GW97} the last equation was replaced by 
$\tau(\gamma\eta)_r=A(Phot) | T_{\eta\eta}|,$
where $A(Phot)$ was 
treated as a free parameter in the Minuit minimization.

At first, because of the lack of uniqueness in the two analyses 
published in  Refs.~\cite{gpivip} and \cite{Mainz},
 only the 32 $S11(\gamma N\rightarrow \pi N)$ amplitudes with 
$E_{c.m.} \leq 1550$ MeV were used,
since this upper energy limit is about the same as for the 
$\gamma N\rightarrow \eta N$ data. This resulted in a
good fit with parameters qualitatively the same as in Ref.~\cite{GW97} and
also in line with the Particle Data Group\cite{PDT} -- see columns A, D
and PDG in Table~\ref{table1}. In column A, the error bars in the 
$S11(\gamma N\rightarrow \pi N)$ amplitudes of Ref.~\cite{gpivip} have
all been increased to $\pm \sqrt{2}$ -- for the reasons discussed earlier.
In this case, the overall $\chi^2$/dof and the separate $\chi^2$/dpt are
all near unity. However, when in column D the errors are increased to
only $\pm 1/\sqrt{2}$, the $\chi^2$/dpt for the $S11(\gamma N\rightarrow
\pi N)$ amplitudes become significantly larger. Columns B and C show the
corresponding results when the $S11(\gamma N\rightarrow \pi N)$ data
base is increased to include data with $E_{c.m.}$ upto 1650MeV. The fits
are now systematically worse than in column A with the overall 
$\chi^2$/dof increasing from 0.89 to 1.23 in column B. In column C -- the case
with smaller errors and the larger data base -- the fit obtained was quite
poor to such an extent that reasonable errors on the parameters could not be 
extracted. The latter fit, when all 13 parameters were varied
simultaneously, did not give from Minuit a Migrad result  that converged
to sensible parameters. The fit displayed in column C is based on the
parameters of column B, some of which are first Fixed and then Released
and Scanned by Minuit. 
The comparison with the data being fitted is shown in Figs. 1--4
The main conclusions to be drawn from Table~\ref{table1} and these figures
are:\\
1) All four fits to the data are reasonable with cases A and B being
superior.\\
2) The main distinguishing feature between the four fits is the relative
ability to fit the $S11(\gamma N\rightarrow \pi N)$ data, since this is
the channel that contributes most to the overall $\chi^2$/dof -- with the
$\chi^2$/dpt's from the other four channels being reasonably constant and
comparable to unity in all fits. This suggests that it will be hard for the
present type of analysis to maintain these latter $\chi^2$/dpt's and, at
the same time, achieve a good  $\chi^2$/dpt for the 
$S11(\gamma N\rightarrow \pi N)$ data presented in Refs.~\cite{gpivip}
and \cite{Mainz}. The authors, therefore, suggest that the $S11(\gamma
N\rightarrow \pi N)$ amplitudes from the $K-$ matrix  model could be a 
more realistic  set than those in Refs.~\cite{gpivip}    
and \cite{Mainz}, since they are now consistent with more reactions 
$\pi  N\rightarrow \pi  N$, $\pi  N\rightarrow \eta N$ and
$\gamma N\rightarrow \eta N$.\\
3) Figure 3 shows that, beyond $E_{c.m.}\approx 1550$MeV, cases A and D
give larger cross sections than B and C -- the difference increasing to
about a factor of two by $E_{c.m.}\approx 1650$MeV. In the near future,
the GRAAL collaboration \cite{GRAAL} is expected to provide total cross
section data upto this energy and so, hopefully, distinguish between
these cases.

In Table~\ref{table1} the parameters $\Gamma(Total)$, $\eta(br)$
$\pi(br)$, $\Gamma(Total,1)$ and $\pi(br,1)$ are quoted, whereas the earlier
formalism is expressed in terms of $\gamma_{\eta}(0)$,
$\gamma_{\pi}(0, \ 1)$, and $\gamma_{3}(0, \ 1)$. 
The two notations are related as follows:\\
1) $\gamma_{\eta}(0)=0.5 \Gamma(Total)\eta(br)/q_{\eta}[E_0(R)]$, 
 \ \ 2) $ \ \ \gamma_{\pi}(0)=0.5 \Gamma(Total)\pi(br)/q_{\pi}[E_0(R)]$,\\ 
3) $\gamma_{\pi}(1)=0.5 \Gamma(Total, \ 1)\pi(br, \ 1)/q_{\pi}[E_1(R)]$,
4) $ \ \ \gamma_{3}(0)=0.5 \Gamma(Total)[1-\eta(br)-\pi(br)]/q_{3}[E_0(R)]$,\\
5) $\gamma_{3}(1)=0.5 \Gamma(Total, \ 1)[1-\pi(br, \ 1)]/q_{3}[E_1(R)]$ and
6) $\Gamma_{\gamma}(0,1)=2 q_{\gamma}[E_{0,1}(R)]\gamma_{\gamma}(0,1)$.\\
This now requires a choice to be made for the reference energies
$E_0 (R)$ and $E_1 (R)$, which preferably should be close to the $E_{0,1}$ 
in Table~\ref{table1}. Here, we take simply 
$E_{0,1}(R)=$ 1535,  1650 MeV respectively. This gives
$q_{\eta}[E_0(R)]=0.945, \ \ q_{\pi}[E_0(R)]=2.365, \ \
q_{\pi}[E_1(R)]=2.770, \ \ q_{3}[E_0(R)]=1.067, \ \ q_{3}[E_1(R)]=1.245,
 \ \ q_{\gamma}[E_{0}(R)]=2.436$ and $ q_{\gamma}[E_{1}(R)]=2.829$ fm$^{-1}$.
It should be added that this is not an assumption or an approximation.
It is just setting a scale that is needed when converting from one
notation to the other.  In Table~\ref{table2}, the
$\gamma_{\pi,\eta,3}(0,1)$
are tabulated alongwith $\Gamma_{\gamma}(0,1)$. 

In the above, we have been very explicit in describing the formalism.
Therefore, in principle, the reader should be able to reconstruct all
three $T$-matrices and so determine each of the complex amplitudes
needed in the five processes
$\pi  N\rightarrow \pi  N$, $\pi  N\rightarrow \eta N$,
$\eta  N\rightarrow \eta N$, $\gamma N\rightarrow \pi N$ and
$\gamma N\rightarrow \eta N$. This formalism also enables
these amplitudes to be calculated at unphysical energies. For example,
in the study of possible $\eta$-nucleus quasi-bound states, the    
$\eta  N\rightarrow \eta N$ amplitudes are needed below the $\eta$
threshold. This is easily achieved by simply using an $\eta$ 
momentum($q_{\eta}$) that is purely imaginary.

In spite of the model being very explicit, it is sometimes convenient
to have simplified versions of some of the amplitudes. The ones
we consider are those that are expansions in terms of  $q_{\eta}$ about 
the $\eta$ threshold -- in particular $A_{\eta \eta}$ and 
$A_{\gamma \eta}$. The former results in the usual 
$\eta  N\rightarrow \eta N$ effective range expansion of
Ref.~\cite{GW97}, the parameters of which are now updated in 
Table~\ref{table3}. 
This shows that the scattering length($a$) is larger than that extracted
in Ref.~\cite{GW97} -- the increase being 15\% for case A and 40\%  for case
B. However, it should be remembered that case B extrapolates the model
into a region where the $\gamma N\rightarrow \eta N$ data is lacking,
and it is just this reaction that is crucial in determining the 
scattering length.
Given this expansion, then $T_{\eta \eta}$
is readily calculated from $A_{\eta \eta}$ as in Eq.~\ref{KT1} at 
energies both above and below the $\eta$ threshold. The other amplitude
of interest is $T_{\gamma \eta}$ in Eq.~\ref{KT2}, which is seen to
depend on both $A_{\eta \eta}$ and $A_{\gamma \eta}$. By analogy with
the expansion of $T_{\eta \eta}$, we express $T_{\gamma \eta}$ in the
form
\begin{equation}
\label{intge}
\frac{1}{T_{\gamma \eta}}=\frac{1}{A_{\gamma \eta}}
-iq_{\eta}\frac{A_{\eta \eta}}{A_{\gamma \eta}}.
\end{equation}
The two entities $1/A_{\gamma \eta}$ and $A_{\eta \eta}/A_{\gamma \eta}$
are then expanded as $e_i+f_iq_{\eta}^2+g_iq_{\eta}^4$ with the
parameters $e_i, \ f_i, \ g_i$ being given in Table~\ref{table4}.
Both of these expansions do very well over the energy range of
Ref.~\cite{Krusche}. For example, with case A at $E_{c.m}=1538.6$MeV --
an energy that is 50 MeV above the $\eta$ threshold -- ,
the expansion of $1/A_{\gamma \eta}$ gives 8.4--i20.0 fm$^{-1}$ compared
with the exact value of 7.9--i19.7 fm$^{-1}$ and the expansion of
$A_{\eta \eta}/A_{\gamma \eta}$ gives 37.12--i5.4 compared
with the exact value of 37.08--i5.7. This latter agreement 
and the weak energy dependence of this quantity explains
why, in Ref.~\cite{GW97}, the replacement of $\sigma (\gamma N\rightarrow
\eta N) \propto |T_{\gamma\eta}|^2$ by $\sigma (\gamma N\rightarrow \eta
N) \propto |T_{\eta\eta}|^2$ was a good approximation, since --
as seen from Eq.~\ref{KT1} -- 
$T_{\gamma\eta}=A_{\gamma\eta}T_{ \eta\eta}/A_{\eta \eta}\rightarrow
A(Phot)T_{ \eta\eta}$. Also the value of $A(Phot)=19.74(36)$ in 
Table~\ref{table1} is essentially given by 
$1/ e_2 \approx \frac {1}{40}$$m_{\pi} \ 10^3 \approx 18$. 
In Fig.~5, the real and imaginary components of $T_{\gamma \eta}$ 
are shown, when these two expansions are used in Eq.\ref{intge}, which is
then inverted to give $T_{\gamma \eta}$. It is
seen that they give a good representation over a wide energy range
especially for energies below the $\eta$ threshold.
This agreement is very similar to that found in Ref.~\cite{GW97} for 
$T_{\eta \eta}$. It should be added that, if the form of 
$T_{\gamma \eta}$ written in Eq.~\ref{KT2} is used directly with 
expansions of $A_{\gamma\eta }$ and $A_{\eta \eta}$, then the fit is
much poorer -- as also seen in Fig.~5.

\section{Discussion and conclusions}
In this paper the authors have developed a simple $K$-matrix parametrization
that gives, in an energy range of about 100 MeV each side of the $\eta$ 
threshold, a good fit to $\pi  N\rightarrow \pi  N$, $\pi  N\rightarrow
\eta N$ and $\gamma N\rightarrow \eta N$ data. In addition, it has the
same trends as the $\gamma N\rightarrow \pi N$ data, which at present
is not unique over this energy range. However, this  consistent fit
should not be considered an end in itself, since it also results in
predictions for the  $\eta  N\rightarrow \eta  N$ $S$-wave amplitude.
Near the $\eta$ threshold this amplitude has been parametrized in the
form of the effective range expansion -- the resultant parameters being
given in Table~\ref{table3}. Since this expansion is good over a wide
energy range each side of the $\eta$ threshold, it is very useful for
discussions concerning the possibility of $\eta$-nucleus quasi-bound states
e.g. in Ref.~\cite{Vldov} the effective range expansion of
Ref.~\cite{GW97} was used to study the production of $\eta$-nuclei, while
Ref.~\cite{Others} uses such an expansion to describe $\eta$-nucleus 
final state interactions. The indications from Table~\ref{table3} are
that the $\eta - N$ scattering length is now larger than that extracted
in Ref.~\cite{GW97}. This is even more favourable for the existence of 
$\eta$-nucleus quasi-bound states and
 may lead to an early onset of nuclear P-wave states, which are 
easier to detect in the Darmstadt experiment outlined in
Ref.~\cite{Hayano}.

   One result of the above fits is the extraction of the 
photon-nucleon-$N$(1535)
coupling constant $\gamma_{\gamma} (0,1)$ as indicated in Table 1, which
 is equivalent to the partial decay width $\Gamma _{\gamma} $
for $N(1535)\rightarrow  \gamma N$. The definition of  $\Gamma _{\gamma} $ is
not unique, however. Below,  this question is elucidated on a simple 
soluble model of the $T$ matrix. This model is also used  to understand
the interference of a resonant interaction  described by a 
singularity in the $K$ matrix and  potential interactions described by the 
background parameters $B$.

Let us, assume a separable $K$ matrix model with
\begin{equation}
\label{sepk}
K_{i,j}= \sqrt{\gamma_{i}\gamma_{j}} (\frac{1}{E_0-E}+B ),
\end{equation}
where, in the notation of Eqs.~\ref{Ks} and \ref{ketagamma}, 
$B_{ij}=B \sqrt{\gamma_{i}\gamma_{j}}$.
This leads to a separable solution for the $T$-matrix
\begin{equation}
\label{sept}
 T_{i,j}= \sqrt{\gamma_{i}\gamma_{j} }
\frac{1+B(E_0-E)}{E_0-E  - i\sum_k q_k\gamma_{k} [1+ B(E_0-E)] }.
\end{equation}
When the background term $B$ vanishes, this model is equivalent to simple
Breit-Wigner multichannel resonances of eigen-width 
$\Gamma/2 = \sum_k q_k\gamma_{k}$. However,
when we relax this restriction  a new structure is built upon the
resonance. It is determined by the energy dependent term  $[1+B(E_0-E)],$
which generates a zero of the cross section at  $E= E_0 + 1/B $.  Now,
it is the $1/B$ that sets a new energy scale, which may be independent
of the scale given by the width. For a large $B$ one finds the resonance
to be accompanied by a nearby zero, whereas  for small $ B$ this zero is
moved away beyond the resonance width.
The resonance shape is thus very different from the Lorentian : 
one reason being the strong energy dependence in 
$ q_{k}(E)$ and another being  the pole-background interference.  

As discussed above in connection with Table~\ref{table2}, it is usually 
natural to 
define the partial width of a resonance on the basis of  Eqs.~\ref{sepk} and 
\ref{sept} as  
$\Gamma _{\gamma}(0,1) = 2 q_{\gamma}(E_{0,1})\gamma_{\gamma}(0,1)$.
For the best 
fits to the data (sets $A, \ \ B$ of Table 1) this equation produces 
$\Gamma _{\gamma}(0)=0.171, \ \ 0.157$ MeV and 
$\Gamma_{\gamma}(1)= 0.0, \ \ 0.080$ MeV respectively .   
However, with complicated phenomenological $T$ matrices one could define 
$\Gamma _{\gamma}$ otherwise, e.g. by moulding $T$ into the Breit-Wigner 
form in the proximity of $E=E_0$.   
Thus, at  $ Re T_{\gamma,j}=0 $ one has  
$ Im T_{\gamma,j} = 
\frac {\sqrt{(\Gamma_{\gamma}/q_{\gamma})(\Gamma_{j}/q_j)}}{ \Gamma }
=\frac {\sqrt{(\Gamma_{\gamma}/2 q_{\gamma}) \gamma_{j}}}{ \Gamma /2 }$.
Inserting from Table~\ref{table2} the values of $\Gamma$ and
$\gamma_{j}$ gives another estimate of $\Gamma_{\gamma}$. For example,
with case $A$, $ Re T_{\gamma,\eta}=0 $ at $E =1540$ MeV giving 
$ Im T_{\gamma,\eta}=0.0179$ and $q_{\gamma}=2.45$ fm$^{-1}$. This
results in $\Gamma_{\gamma}=0.176$ MeV.
Similarly, $ Re T_{\gamma,\pi}=0 $ at $E =1535$ MeV giving
$ Im T_{\gamma,\pi}=0.0085$ and $q_{\gamma}=2.43$ fm$^{-1}$. This
results in $\Gamma_{\gamma}=0.150$ MeV.
The proximity of these three widths reflects the fact that the realistic
situation is fairly  close to the separability situation described by 
Eqs.~\ref{sepk} and \ref{sept}. It is found to hold approximately,  
for all the parameter sets in Table 1.

   There is another, somewhat unexpected effect of the $[1+B(E_0-E)]$
interference term in Eq. \ref{sept}. For $B>0$ one finds that the 
amplitudes below the resonance are enhanced, and the amplitudes above
the resonance are reduced with respect to the pure resonance term. 
This effect is seen clearly for the $B$ and $C$ parameter sets, where 
below the resonance the 
 $B_{\eta \eta}$ parameter in the  $\eta-N$ channel  is the largest and the
real parts of  the $\eta-N$ scattering lengths given in Table~\ref{table3}  
are also the largest. On the other hand the  $(\gamma,\eta)$ production 
cross section, dominated by the final state  $\eta-N$ interactions, 
become the smallest above the resonance as seen Fig.~4.
One consequence  of this effect 
is that an  extension of the $(\gamma,\eta)$ cross section  measurements 
to energies above the $N(1535)$ resonance may by instrumental in fixing 
more precisely the 
$\eta-N$ scattering length. As indicated in the introduction, the real 
part of this scattering length is crucial in the determination of 
quasibound states in  $\eta$--few nucleon systems. 

On the experimental side there are several
groups \cite{Krusche},\cite{GRAAL}  studying the
$\gamma N\rightarrow \eta N$ reaction in or near this interesting 
energy range.
The observation of 
the cross section   near and above $E_{c.m.}$=1540MeV
would be of great interest enabling a detailed study to be made of the 
$N(1535)$ and possibly leading to a better understanding of the 
internal structure of this object. At present there is no definite
conclusion as to whether or not this resonance structure is due to a
pole in the $K$-matrix, as advocated here, or arising through coupling to
high lying closed channels - see Ref.~\cite{AGWW}.

In the near future, the authors of Ref.~\cite{gpivip} are expected to
extract, directly from experiment, separate values for the real and imaginary
components of $T(\gamma \eta)$. These will be analogous to the 
$S11\gamma N\rightarrow \pi N$ data already available in
Ref.~\cite{gpivip} and used in the above fits. 
Such a development will then enable the
present type of $K$-matrix analysis to be even more constrained.

One of the authors (S.W.) wishes to acknowledge the hospitality of the
Research Institute for Theoretical Physics, Helsinki, where part of this 
work was carried out. 
In addition he was partially supported by grant  No  KBN 2P03B 016 15.
The authors also thank Drs. R. Arndt and B. Krusche
for useful correspondence and members of the GRAAL collaboration for
useful discussions.
This line of research involving $\eta$-mesons is partially supported by
the Academy of Finland.

%\newpage

%\newpage
%\clearpage

\section*{}

\begin{table}
\begin{center}
\caption{ The optimised parameters from Minuit defining the $K$-matrices:
There are in column A only 32 $\gamma N\rightarrow \pi N$ data points
 with $E_{cm}\le 1550$ MeV and error bars all 1.41, 
in column B 48 data points with $E_{cm}\le 1650$ MeV 
and error bars all 1.41. Cases D and C are the same as A and B except
that the error bars are reduced to 0.70.
In addition, the first column shows the results from
Ref.~\protect\cite{GW97}
and  the last column  the  corresponding values from the
Particle Data Group \protect\cite{PDT}.}
\vspace{1cm} 
\begin{tabular}{ccccccc}  
&\protect\cite{GW97}&A&D&B&C&\protect\cite{PDT} \\ \hline
$B_{\eta \eta}$(fm)&0.177(33)&0.263(32)&0.228(106)&0.371(48)&0.372&--\\
$B_{\pi \eta}$(fm)&0.022(13)&0.016(8)&0.027(17)&0.003(15)&0.019&--\\
$E_0$(MeV)&1541.0(1.6)&1536.8(0.9)&1540.6(6.6)&1530.0(2.5)&1529.5&1535(20)\\
$E_1$(MeV)&1681.6(1.6)&1682.1(1.6)&1683.2(1.6)&1682.9(1.6)&1685.4&1650(30)\\
$\Gamma(Total)$(MeV)&148.2(8.1) &138.2(1.3)&142.7(13.4)&122.4(5.0)&114.4
&100--250\\
$\eta(br)$&0.568(11)&0.585(8)&0.594(18)&0.61(17)&0.648&0.30--0.55 \\ 
$\pi(br)$&0.394(9)&0.380(6)&0.371(13)&0.358(8)&0.330&0.35--0.55 \\ 
$\Gamma(Total,1)$(MeV)&167.9(9.4)&171.7(6.3)&183.8(9.2)&178.8(7.8)&203.8&145--190 \\ 
$\pi(br,1)$&0.735(11)&0.729(10)&0.721(11)&0.724(10)&0.709&0.55--0.90 \\ 
$A(Phot)$&19.74(36)&--&--&--&--&-- \\ 
$B_{\gamma \eta}$(fm)&--&0.0040(1)&0.0049(15)&0.0027(9)&0.0036&-- \\
$B_{\gamma \pi}$(fm)&--&0.0030(4)&0.0034(5)&0.0013(7)&0.0021&-- \\
$\gamma_{\gamma}(0)$& --&0.00018(1)&0.00018(2)&0.00016(1)&0.00014&-- \\
$\gamma_{\gamma}(1)$& --&$\le 10^{-8}$
&$\le 10^{-8}$&7(2)$10^{-5}$&6(1)$10^{-5}$&-- \\  \hline
$\chi^2(\gamma N\rightarrow \eta N)$/dpt&0.75&0.75&0.75&0.79&0.94&\\
$\chi^2(\pi  N\rightarrow \eta N)$/dpt&0.73&0.69&0.67&0.93&2.10&\\
$\chi^2(\pi  N\rightarrow \pi N)$/dpt $\cal R$&0.94&1.04&1.52&1.27&2.89&\\
$\chi^2(\pi  N\rightarrow \pi N)$/dpt $\cal I$&0.60&0.56&0.63&0.48&0.92&\\
$\chi^2(\gamma N\rightarrow \pi N)$/dpt $\cal R$&--&0.59&2.09&1.34&3.83&\\
$\chi^2(\gamma N\rightarrow \pi N)$/dpt $\cal I$&--&1.33&4.27&2.25&5.53&\\ 
$\chi^2$(Total)/dof&0.83&0.89&1.54&1.23&2.66&\\ \hline
\end{tabular}
\label{table1}
\end{center}

\vskip 1.0 cm
\end{table}

\newpage

\begin{table}
\begin{center}
\caption{Conversion from the parameters  $\Gamma(Total)$, $\eta(br)$
$\pi(br)$, $\Gamma(Total,1)$,  $\pi(br,1)$ in Table~\protect\ref{table1}
and to the parameters $\gamma_{\eta}(0)$,
$\gamma_{\pi}(0, \ 1)$, and $\gamma_{3}(0, \ 1)$ used in the formalism.
 The conversion $\gamma_{\gamma}(0,1)$ to $\Gamma_{\gamma}(0,1)$ and the
2-$\pi$ branching ratio(2-$\pi$ br) are also given. The other notations
are the same as Table~1. }

\vspace{1cm} 
\begin{tabular}{cccccc}  
&\protect\cite{GW97}&A&D&B&C \\ \hline
$\gamma_{\eta}(0)$&0.226&0.217&0.227&0.202&0.199\\
$\gamma_{\pi}(0)$&0.063&0.056&0.057&0.047&0.040\\
$\gamma_{\pi}(1)$&0.113&0.115&0.121&0.118&0.132\\
2-$\pi$ br&0.038&0.035&0.035&0.027&0.022\\
$\gamma_{3}(0)$&0.0134&0.0114&0.0120&0.0079&0.0059\\
$\gamma_{3}(1)$&0.0906&0.0945&0.1043&0.1004&0.1208\\
$\Gamma_{\gamma}(0)$(MeV)&--&0.171&0.172&0.157&0.136\\
$\Gamma_{\gamma}(1)$(MeV)&--&0&0&0.080&0.068\\
\\ \hline
\end{tabular}
\label{table2}
\end{center}
\vskip 1.0 cm
\end{table}

\newpage

\begin{table}
\begin{center}
\caption{ Results for the scattering length($a$), effective range($r_0$)
and Shape parameter($s$) compared with earlier works. The  other notations
are the same as Table~1.  }
\vspace{0.5cm} 
\begin{tabular}{cccccc} 
&\protect\cite{GW97}&A&D&B&C \\ \hline
$a$ in fm& 0.75(4)+i0.27(3)&0.87+i0.27&0.83+i0.27
&1.05+i0.27&1.07+i0.26\\ 
$r_0$ in fm&--1.50(13)--i0.24(4)&--1.31--i0.28
&--1.34--i0.22&--1.19--i0.31&--1.25--i0.25\\
$s$ in fm$^3$&--0.10(2)--i0.01(1)&--0.14--i0.03
&--0.12--i0.01&--0.18--i0.06&--0.20--i0.05 \\ \hline
\end{tabular}
\label{table3}
\end{center}
\end{table}

\newpage

\begin{table}
\begin{center}
\caption{The parameters $e_i, \ f_i, \ g_i$ in the expansions 
$e_i+f_iq_{\eta}^2+g_iq_{\eta}^4$ for $1/A_{\gamma \eta}$($i=1$) and 
$A_{\eta\eta}/A_{\gamma\eta}$($i=2$). The  other notations
are the same as Table~1.   }
\vspace{0.5cm} 
\begin{tabular}{ccccc} 
For $1/A_{\gamma \eta}$&A&D&B&C \\ \hline
$e_1$ in fm$^{-1}$&40.7--i18.6&40.9--i18.0&39.9--i20.5&40.6--i19.3\\
$f_1$ in fm&--31.0--i2.0&--27.7--i1.9&--39.4--i0.4&--39.1+i0.2\\
$g_1$ in fm$^3$&--3.9+i0.6&--4.1+i0.6&--4.1+i2.0&--6.2+i2.2\\ \hline
&&&&\\
For $A_{\eta\eta}/A_{\gamma\eta}$&&&&\\ \hline
$e_2$ &40.6--i5.2&38.6--i4.0&47.4--i10.9&48.4--i10.2\\
$f_2$ in fm$^2$&--3.4--i0.7&--1.1--i0.5&--10.1--i2.1&--8.2--i2.0\\
$g_2$ in fm$^4$&--0.4+i0.1&--0.1+i0.1&--0.8+i0.1&--1.1+i0.1\\ \hline
\hline
\end{tabular}
\label{table4}
\end{center}
\end{table}

%\end{document}

\newpage

\clearpage

\section*{}

\begin{figure}[ht]
%\special{psfile=vipppr.ps hoffset=-35 voffset=-600 hscale=75
\includegraphics{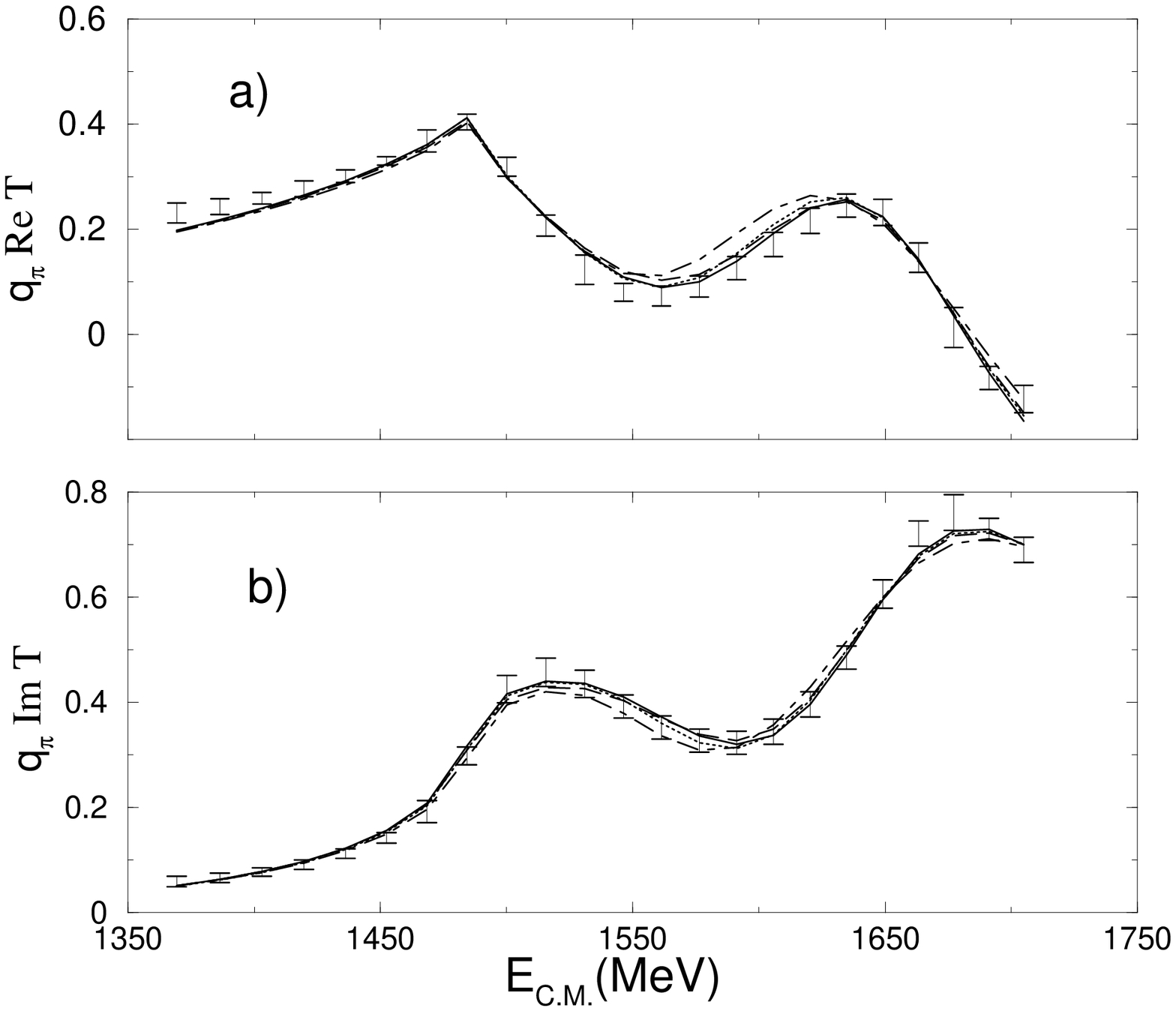}
\caption{$\pi N\rightarrow \pi N$ amplitudes Real and Imaginary}
\label{fig.1r}
\end{figure}

\newpage

\begin{figure}[ht]
%\special{psfile=nefabcd.ps hoffset=-35 voffset=-600 hscale=75
\includegraphics{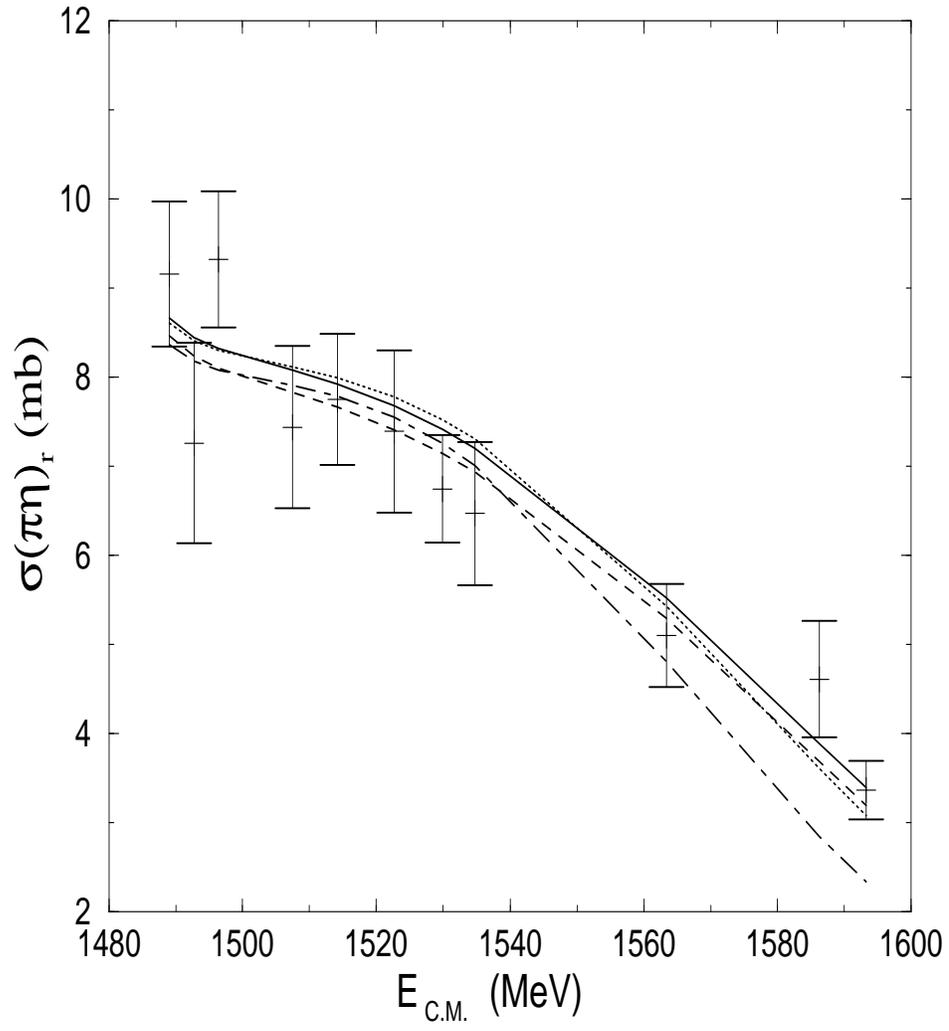}
\caption{Nefkens data}
\label{fig.3}
\end{figure}

\newpage

\begin{figure}[ht]
%\special{psfile= krusabcdl.ps hoffset=-35 voffset=-600 hscale=75
\includegraphics{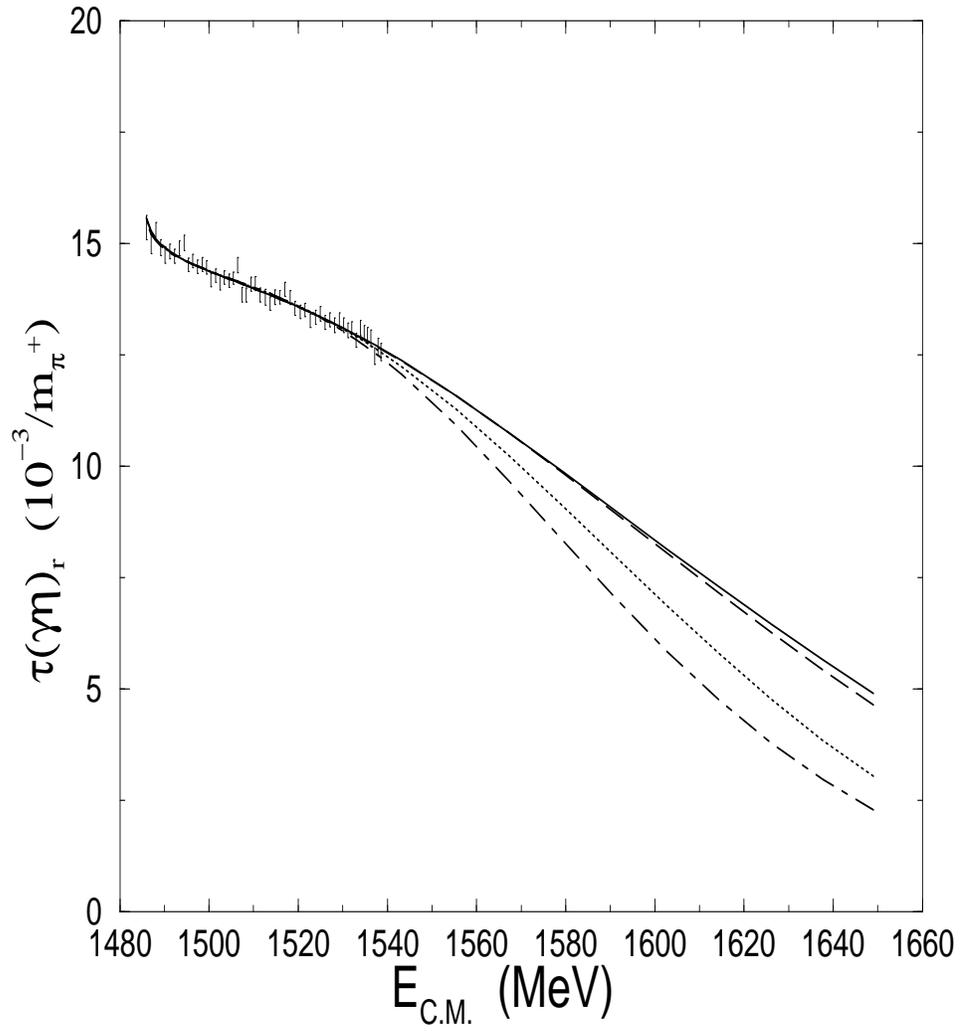}
\caption{Krusche extrapolation}
\label{fig.4a}
\end{figure}

\newpage

\begin{figure}[ht]
%\special{psfile=gprabcd.ps hoffset=-35 voffset=-600 hscale=75
\includegraphics{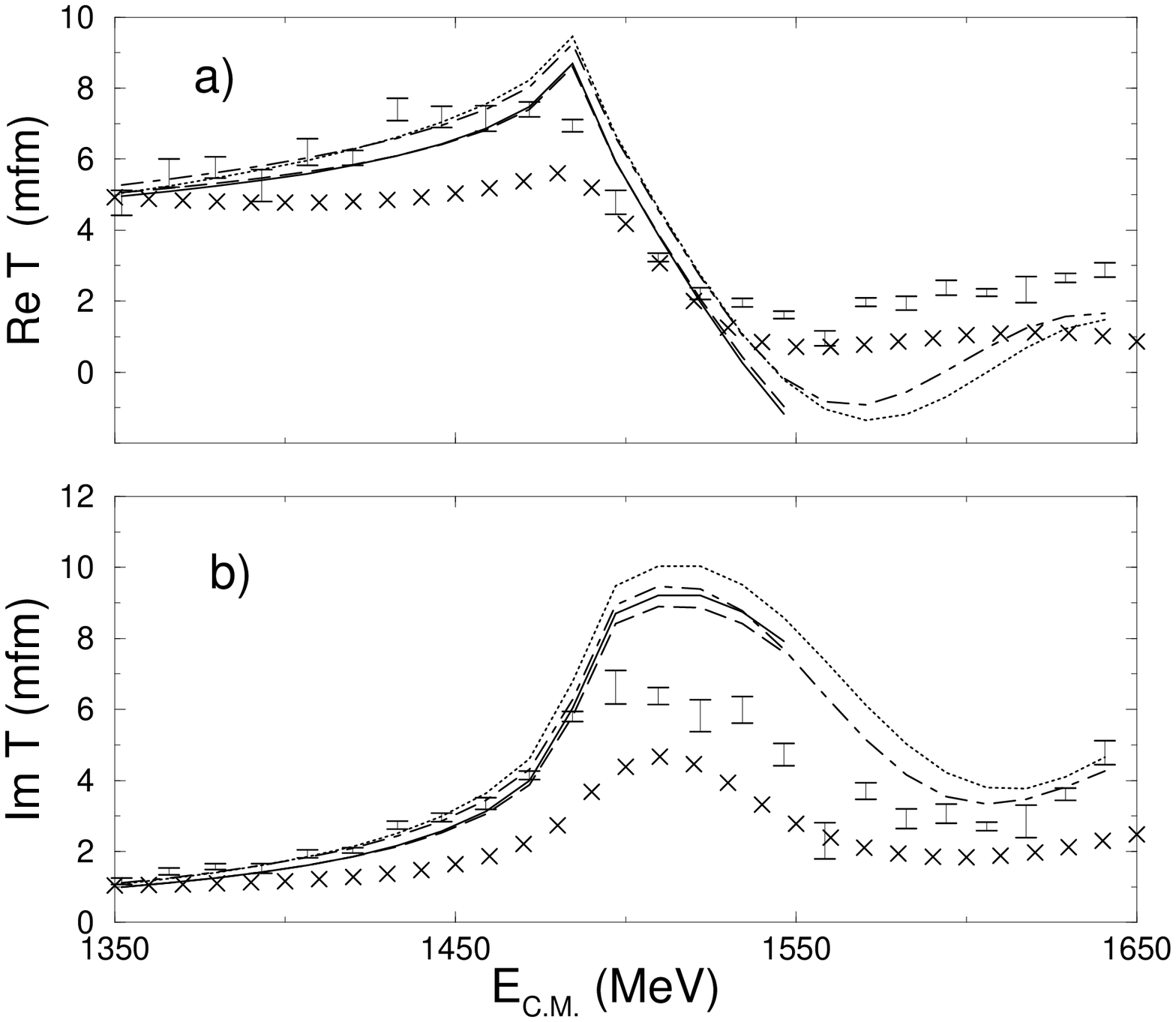}
\caption{$\gamma N\rightarrow \pi N$ amplitudes Real and Imaginary}
\label{fig.5a}
\end{figure}

\newpage

\begin{figure}[ht]
%\special{psfile=tger.ps hoffset=-35 voffset=-600 hscale=75
\includegraphics{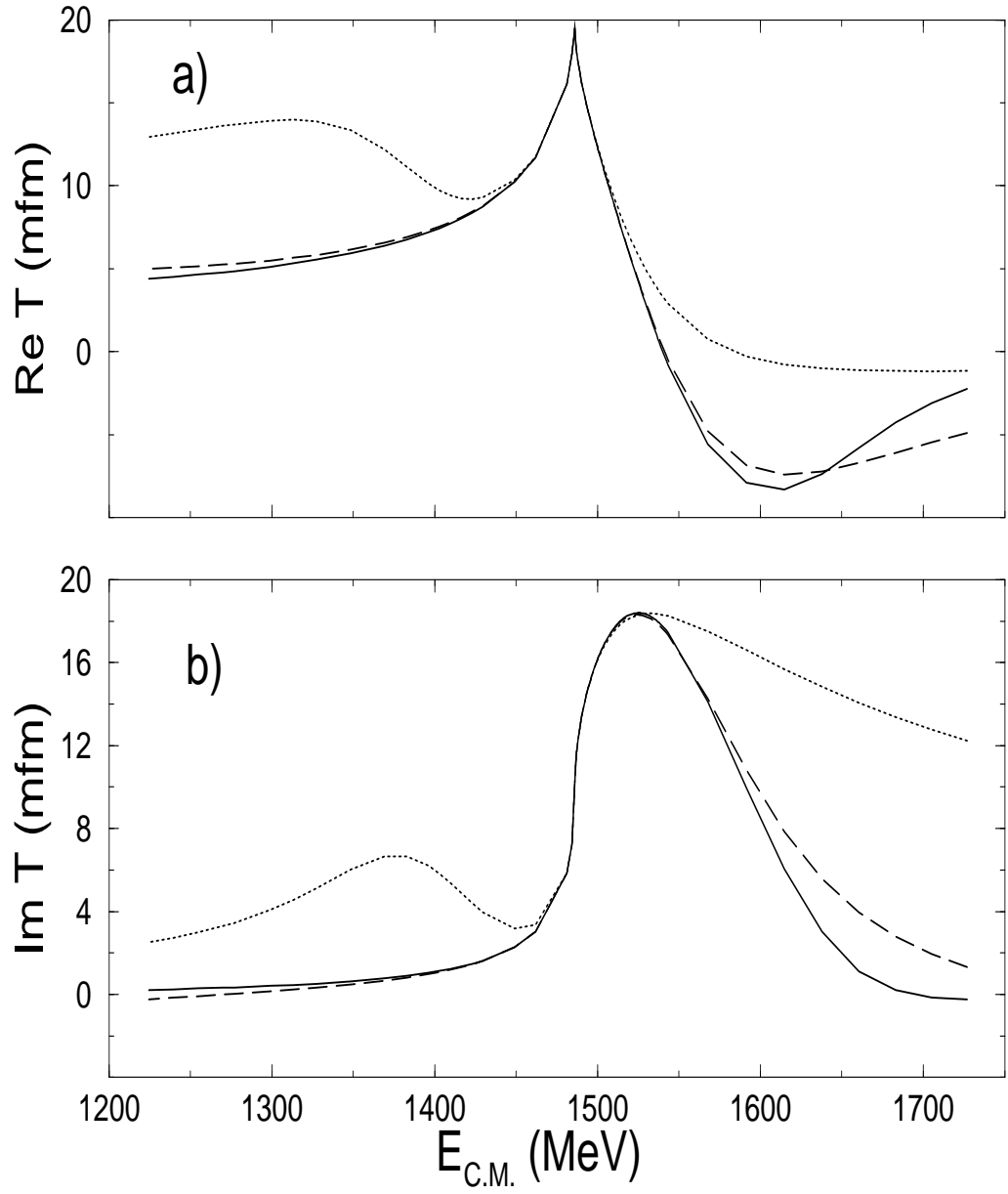}
\caption{$T(\gamma,\eta)$ amplitudes Real and Imaginary}
\label{fig.7}
\end{figure}

\newpage

\begin{center}
{\bf Figure Captions}
\end{center}
\vspace{0.5cm} 

\noindent Figure 1.  The a) Real and b) Imaginary parts of the 
s-wave $\pi  N\rightarrow \pi  N$ amplitudes of Ref.~\protect\cite{Arndt}.
Solid line for solution A,
dashed for D, dotted for  B and dash-dot for  C.

\vspace{0.5cm} 

\noindent Figure 2.  The $\pi  N\rightarrow \eta N$ reaction. Data is  from
Ref.~\protect\cite{Nefkens}. Notation as in Figure 1.

\vspace{0.5cm} 

\noindent Figure 3.  The $\gamma N\rightarrow \eta N$ reaction. Data is from
Ref.~\protect\cite{Krusche}.  Notation as in Figure 1.

\vspace{0.5cm} 

\noindent Figure 4.  The a) Real and b) Imaginary parts of the 
$S11\gamma N\rightarrow \pi N$ of
Ref.~\protect\cite{gpivip}. Crosses are data from Ref.~\protect\cite{Mainz}.
  Notation as in Figure 1.

\vspace{0.5cm}

\noindent Figure 5.  The a) Real and b) Imaginary parts of $T_{\gamma \eta}$.
The solid curve is the exact value as given by the model. 
The dotted curve uses expansions of $1/A_{\gamma \eta}$ and 
$A_{\eta \eta}/A_{\gamma \eta}$, whereas the dashed curve uses those 
for $A_{\gamma \eta}$ and $A_{\eta\eta}$.


\begin{thebibliography}{99}
\bibitem{Krusche}  B. Krusche et al., Phys.Rev.Lett. {\bf 74},  3736 (1995)\\
M. Fuchs et al., Phys.Lett. {\bf B368}, 20 (1996).

\bibitem{GRAAL}  J. Ajaka et al., Phys.Rev.Lett. {\bf 81},  1797 (1998).

\bibitem{saturne} E. Chiavassa et al., Phys.Lett. {\bf B337},  192 (1994).

\bibitem{uppsala} H. Cal\'{e}n et al., Phys.Lett. {\bf B366},  39
(1996).\\
A.M. Bergdold et al. Phys.  Rev. D{\bf 48}, R2968 (1993)

\bibitem{hadronic} 
 J.T Balewski et al., Acta Phys. Pol. {\bf B27}, 2911
(1996).

\bibitem{bha} 
R.S. Bhalerao and L.C.Liu,
Phys.Rev. Lett. {\bf 54}, 286 (1985). 

\bibitem{hai} 
Q.Haider and L.C.Liu,
Phys. Lett. {\bf B172}, 257 (1986); Phys. Rev. C{\bf 34}, 1845(1986) . 

\bibitem{sau} Ch. Sauerman, B.L. Friman and W. N\"orenberg,
Phys. Lett. {\bf B341}, 261  (1995).

\bibitem{ben} C. Benhold and H. Tanabe,
Nucl. Phys. {\bf A350}, 625  (1991).

\bibitem{kai} N. Kaiser,  P.B. Siegel and W. Weise,
Nucl.Phys. Lett.{\bf A594}, 325  (1995).

\bibitem{ari} M. Arima, K. Shimizu and K. Yazaki,
Nucl. Phys. {\bf A543}, 613  (1992).

\bibitem{GW97} A. M. Green and S. Wycech, Phys. Rev. C{\bf 55}, R2167 (1997)
\bibitem{watson} M.L. Goldberger and K.M. Watson, {\em Collision Theory}
( Wiley, New York -- London, 1964)
\bibitem{Arndt}R.Arndt, J.M.Ford and L.D.Roper, Phys.Rev. D{\bf 32}, 
1085 (1985).

Solution SM95 obtained from SAID via Internet in November 1996.
\bibitem{Nefkens} M.Clajus and B.M.K.Nefkens , $\pi$-N newsletter $\bf 7$,
76 (1992).

\bibitem{gpivip} R. A. Arndt, I. Strakovski and R. L. Workman,
 Phys. Rev. C{\bf 53}, 430 (1996) and http://said.phys.vt.edu/

\bibitem{Mainz}D. Drechsel, O. Hanstein, S.S. Kamalov and L. Tiator,
Nucl. Phys. {\bf A645},145(1999)


\bibitem{PDT} Particle Data Group,  The European Phys. Journal {\bf C3},
1 (1998) 


\bibitem{Vldov} A.I. L'vov, "Production and decay of $\eta-$mesic nuclei",
Proc.Int.Conf. on Mesons and Nuclei, Prudhonice, Prague, 1998. 
Ed. J. Adams (World Scientific): nucl-th/9809054

G.A. Sokol et al., "Search for $\eta-$mesic nuclei in photo-mesonic
processes", nucl-ex/9812007

\bibitem{Others}
M.Effenberger and A. Sibirtsev , Nucl. Phys. {\bf A632}, 99 (1998)

\bibitem{Hayano}
R.S.Hayano, S.Hirenzaki and A.Gillitzer , 
"Formation of $\eta$-mesic Nuclei Using the Recoilless (d,$^3$He)
Reaction", nucl-th/9806012.

\bibitem{AGWW} R. A. Arndt, A.M. Green, R. L. Workman and S. Wycech,
Phys. Rev. {\bf C58}, 3636 (1998)


\end{thebibliography}
\end{document}